\documentclass[iop, apj, twocolappendix, numberedappendix, appendixfloats]{emulateapj}
\usepackage{threeparttable}
\usepackage{multirow}
\usepackage{times}
\usepackage{enumitem}
\usepackage{url} 
\usepackage{color} 
\usepackage{amsmath,bm}

\shorttitle{}
\shortauthors{Y. Huang et al.}

\begin{document}

\title{Member stars of the GD-1 tidal stream from the SDSS, LAMOST and Gaia surveys}

\author{Y. Huang\altaffilmark{1,7}}
\author{ B.-Q. Chen\altaffilmark{1}}
\author{ H.-W. Zhang\altaffilmark{2,3}}
\author{ H.-B. Yuan\altaffilmark{4}}
\author{ M.-S. Xiang\altaffilmark{5}}
\author{ C. Wang\altaffilmark{2,6}}
\author{ Z.-J. Tian\altaffilmark{1}}
\author{ X.-W. Liu\altaffilmark{1,7}}

\altaffiltext{1}{South-Western Institute for Astronomy Research, Yunnan University, Kunming 650500, People's Republic of China; {\it yanghuang@ynu.edu.cn {\rm (YH)}; x.liu@ynu.edu.cn {\rm (XWL)}}}
\altaffiltext{2}{Department of Astronomy, Peking University, Beijing 100871, People's Republic of China}
\altaffiltext{3}{Kavli Institute for Astronomy and Astrophysics, Peking University, Beijing 100871, People's Republic of China}
\altaffiltext{4}{Department of Astronomy, Beijing Normal University, Beijing 100875, People's Republic of China}
\altaffiltext{5}{Max-Planck Institute for Astronomy, K{\"o}nigstuhl, D-69117, Heidelberg, Germany}
\altaffiltext{6}{LAMOST Fellow}
\altaffiltext{7}{Y. Huang and X.-W. Liu are equal corresponding authors.}

\begin{abstract}
With the photometric data from the SDSS survey, the spectroscopic data from the SDSS/SEGUE and the LAMOST surveys, and the astrometric data from the Gaia DR2,
we have identified 67 highly-probable member stars of the GD-1 cold stellar stream spread along almost its entire length (i.e. from 126 to 203 degree in Right Ascension).
With the accurate spectroscopic (i.e. metallicity and line-of-sight velocity) and astrometric (i.e. proper motions) information, 
the position-velocity diagrams, i.e. $\phi_{1}$--$\mu_{\alpha}$, $\phi_{1}$--$\mu_{\delta}$ and $\phi_{1}$--${v_{\rm gsr}}$, of the GD-1 stream are well mapped.
The stream has an average metallicity [Fe/H]\,$= -1.96$.
The rich information of member stars of the stream now available allow one not only to model its origin, but also to place strong constraints on the mass distribution and the gravitational potential of the Milky Way.
\end{abstract}

\section{Introduction}
Stellar streams are relics of tidally disrupted dwarf galaxies or gloubular clusters by the gravitational potential of the  accreting host galaxy or cluster.
In the past decade alone, over a dozen stellar streams have been discovered in the Milky Way (e.g. Odenkirchen et al. 2001; Ibata et val. 2001; Newberg et al. 2002; Majewski et al. 2004; Belokurov et al. 2006; Grillmair 2006), largely owing to the large-scale surveys such as SDSS (York et al. 2000), 2MASS (Skrutskie et al. 2006), PanSTARRS-1 (Chambers et al. 2016) and DES  (DES Collaboration 2016).
The discoveries of these stellar streams provide strong evidence for the hierarchical galaxy formation scenario on galaxy scale (Peebles 1965; Press \& Schechter 1974; Blumenthal et al. 1984) and support the standard $\Lambda$CDM  cosmological model (e.g. Diemand et al. 2008). 

In addition to their cosmological significance, stellar streams, especially the thin and cold ones (e.g. the Pal\,5, GD-1 and NGC\,5466 streams; Lux et al. 2013), are also powerful probes of the Galactic mass distribution and gravitational potential (e.g. Koposov, Rix \& Hogg 2010; Law \& Majewski 2010; Gibbons, Belokurov \& Evans 2014; Bowden, Belokurov \& Evans 2015; K{\"u}pper et al. 2015; Bovy et al. 2016).
Amongst the thin and cold streams, the GD-1 stream (Grillmair \& Dionatos 2006) is the thinnest known, with a width less than 0.25$^{\circ}$ (smaller than 50\,pc assuming a typical distance of $8.5$\,kpc) and a length over $80^{\circ}$ across the northern sky.
Using spectroscopic data from the SDSS/SEGUE, Willett et al. (2009) find an average metallicity for the GD-1 stream of [Fe/H] = $-2.1 \pm 0.1$ and an age comparable to the globular cluster M92.
On the other hand, a relatively more metal-rich ([Fe/H] = $-1.4$) and younger age (9 Gyr) for the GD-1 stream are found by Koposov et al. (2010), based on the isochrone fitting to the photometric data.
To date, no apparent progenitor has been identified for this stream.
Nevertheless,  its low velocity dispersion suggests a low-mass globular cluster origin.

The narrow, long, cold nature of the GD-1 stream makes it  an ideal stream to probe the gravitational potential and dark matter distribution in the inner Milky Way.
However, the number of currently known highly-probable member stars of the GD-1 stream is very limited.
Hitherto, almost all the efforts to constrain the Galactic dark matter distribution using the GD-1 stream are based on the level of $\sim$20 member stars that spread over a limited length of the stream,  from $-12^{\circ}$ to $-45^{\circ}$ in $\phi_{1}$ (of the GD-1 coordinate system; Koposov 2010).
More recently, a much larger number of member stars of this stream has been successfully identified by Price-Whelan \& Bonaca (2018), based on the PanSTARRS-1 photometry and the accurate proper motions from the Gaia DR2 (Lindegren et al. 2018).
Nevertheless, there are no spectroscopic information of those newly identified member stars.

In this paper, we attempt to identify more member stars belonging the GD-1 stream across its entire length, based on the photometric data from the SDSS, the spectroscopic data from the SDSS/SEGUE and the LAMOST surveys, and the astrometric data from the Gaia DR2 (Lindegren et al. 2018).
The multi-dimensional information of the stream member stars will allow one not only to constrain its origin, but also to help probe the gravitational potential of the Milky Way.
The paper is structured as follows. 
In Section\,2, we describe the data used in this paper.
We present the member star selection of the GD-1 stream in Section\,3.
We discuss the main results and draw our conclusions in Sections\,4 and 5.

\section{Data}
\subsection{Photometric data}
In this work, we use the photometric data from the twelfth data releases of the SDSS (DR12; Alam et al. 2015).
All the imaging data were obtained between September 19, 1998 and November 18, 2009, resulting a total of around 35,000 square degrees of images that cover a footprint of 14,055 square degrees of the sky.
The data yielded thousand million (hundred million unique) objects with very accurate multi-band photometry.

\subsection{Spectroscopic data}
The spectroscopic data are from the SDSS/SEGUE DR12 (Alam et al. 2012) and the third release of the LSS-GAC value-added catalogues (Huang et al., in preparation).

SDSS/SEGUE is the SDSS-II/III Extension for Galactic Understanding Exploration that yields a total of about 360,000 optical ($\lambda\lambda$3820--9100), low-resolution ($R \sim 2000$) spectra of Galactic stars at distances ranging from 0.5 to 100 kpc (Yanny et al. 2009). 
Stellar atmospheric parameters (effective temperature $T_{\rm eff}$, surface gravity log\,$g$ and metallicity [Fe/H]) and line-of-sight velocity $v_{\rm los}$ are deduced from those spectra with the SEGUE Stellar Parameter Pipeline (SSPP; Allende Prieto et al. 2008; Lee et al. 2008a,b; Smolinski et al. 2011).
Typical uncertainties are 5\,km\,s$^{-1}$ for $v_{\rm los}$, $180$\,K for $T_{\rm eff}$, 0.24\,dex for log\,$g$ and 0.23\,dex for [Fe/H] (Smolinski et al. 2011).
We note that a 7\,km\,s$^{-1}$ systematic error has been corrected for the derived line-of-sight velocities.

LAMOST is a 4-metre, quasi-meridian reflecting Schmidt telescope equipped with 4000 fibers distributed in a field of view of $5^{\circ}$ in diameter (Cui et al. 2012). 
It can simultaneously collect upto 4000 spectra per exposure, with a wavelength coverage and a resolution similar to those of SDSS.
The scientific motivations and target selections of the surveys are described in Zhao et al. (2012), Deng et al. (2012) and Liu et al. (2014). 
The five-year Phase-I LAMOST Regular Surveys were completed in, summer 2017.
The Phase-II LAMOST Pilot Surveys were initiated September 2017, adding a new component of medium resolution ($R \sim 7500$) surveys.
Two stellar parameter pipelines  have been developed --  the LAMOST Stellar Parameter Pipeline (LASP; Luo et al. 2015) and the LAMOST Stellar Parameter at Peking University (LSP3; Xiang et al.2015, 2017), to derive the stellar atmospheric parameters and line-of-sight velocities from the spectra.
Both pipelines achieve typical uncertainties of 5\,km\,s$^{-1}$ in $v_{\rm los}$\footnote{Values of $v_{\rm los}$ yielded by LSP3 have been corrected for a systematic offset of 3.1 km\,s$^{-1}$ (Xiang et al. 2015; Huang et al. 2018a).},  150\,K in $T_{\rm eff}$, 0.25\,dex in log\,$g$ and 0.15\,dex\footnote{For metal-poor or other types of stars, the precision of metallicity estimates is about 0.20-0.30 or larger.} in [Fe/H]  for `normal' metal-rich ([Fe/H]\,$\ge -1.5$) type (FGK) stars of spectra of SNRs greater than 10.
With the latest version of LSP3 (Xiang et al. 2017), we have derived line-of-sight velocities and atmospheric parameters from approximately 6.5 million stellar spectra (of 4.4 million
unique stars) of signal-to-noise ratios (SNRs) higher than 10, collected by June, 2016.
Additional parameters have been derived, including values of the interstellar reddening, absolute magnitudes and distances, as well as elemental abundance ratios ([$\alpha$/Fe], [C/H] and [N/H]).
The whole data set will be publicly available in the third release of the LSS-GAC value-added catalogues (Huang et al., in preparation).

Since the current work makes use of two spectroscopic data sets, one from the SDSS/SEGUE and another from the LAMOST surveys, 
it is important to ensure that parameters derived from the two data sets are on the same scale.
We note that both SDSS/SEGUE and LAMOST  line-of-sight velocities have been calibrated against measurements from high-resolution spectroscopic results.
The LSP3 metallicities have been calibrated by Huang et al. (2018b) against high-resolution spectroscopic results.
Generally, no significant systematic uncertainty are found for metal-rich stars (i.e. [Fe/H] $> -1.4$) while an offset of  $\sim -0.23$\,dex is found for more metal-poor ones.
To tie the SSPP  metallicities to the same scale of LSP3, we compare the results of over 20,000 common stars.
The comparison shows that LSP3 metallicities are on average 0.06\,dex higher than those yielded by SSPP.
After applying the correction, metallicities from both the surveys should be on the same scale.
Using those common stars, we also examine the zero-points of the calibrated line-of-sight velocities derived from LAMOST and SEGUE data, and find no significant systematic offset between them.

\subsection{Astrometric data}
The astrometic data are taken from the recently released Gaia DR2 (Gaia Collaboration et al. 2018; Lindegren et al. 2018).
The release provide positions, parallaxes and proper motions of over 1.3 billion Galactic stars down to $G \sim 20.7$ (Lindegren et al. 2018).
The typical uncertainties of parallaxes are 0.04, 0.1 and 0.7 mas, respectively, for stars of $G < 14$, $= 17$ and $= 20$\,mag.
For the proper motions, typical uncertainties are 0.05, 0.2 and 1.2 mas\,yr$^{-1}$, respectively.
In this work, the $G$ band magnitudes of the potential GD-1 stream member stars range between 14 and 20\,mag, with a median value of about 17.4\,mag.

\section{Stream member star selection}
In this section, we attempt to identify highly-probable member stars of the GD-1 streamfclear using the data described in Section\,2. 

\subsection{Positions and metallicity cuts}
First, we select potential member stars with spectroscopic observations of the GD-1 stream along its track on the sky.

\begin{figure*}
\begin{center}
\includegraphics[scale=0.65,angle=0]{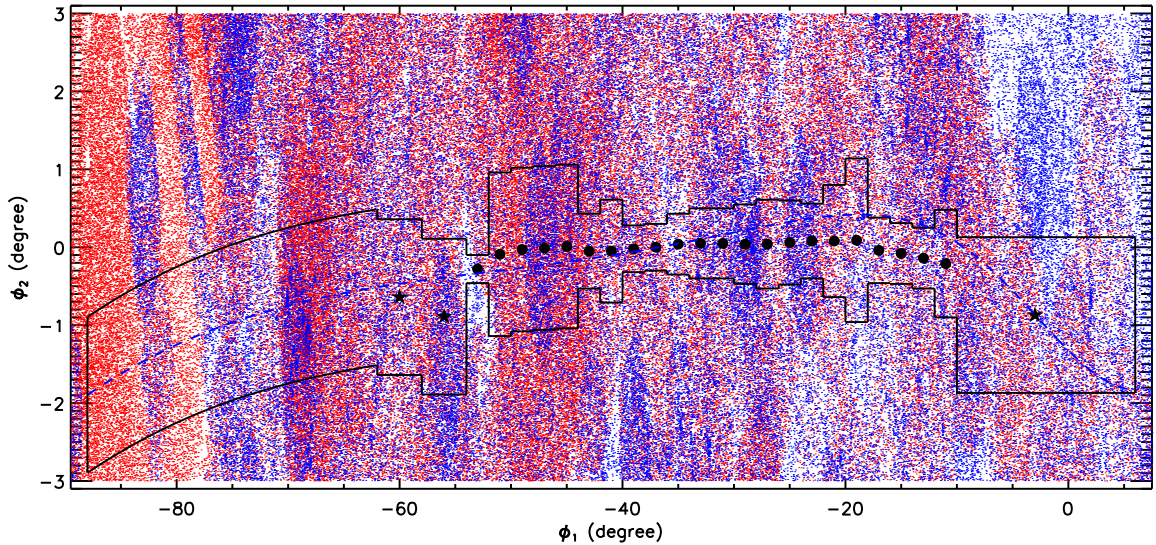}
\caption{Stream position of the GD-1 stream. 
              The black dots, black stars, and blue dash line represent the central positions, derived by B18 and K10 and W09, respectively.
              The black irregular box represents the stream area adopted here for selecting the GD-1 member candidates.
              The background  red and blue dots are the stars observed by the LAMOST and the SDSS/SEGUE surveys, respectively.}
\end{center}
\end{figure*}

The stream was first found by Grillmair \& Dionatos (2006; hereafter GD06).
It is a thin stripe from (RA, Dec) = ($140^{\circ}$, $25^{\circ}$) to (RA, Dec) = ($220^{\circ}$, $55^{\circ}$).
Willett et al. (2009; hereafter W09) use a third-order polynomial to fit the stream track on the sky.
The stream track is redetermined by Koposov et al. (2010; hereafter K10).
They find in stream coordinates ($\phi_{1}$, $\phi_{2}$), the stream extends in $\phi_{1}$ from $-2^{\circ}$ to $-60^{\circ}$.
More recently, using deep photometric data from Megacam mounted on the Canada-France-Hawaii Telescope, de Boer et al. (2018; hereafter B18) derived accurate stream position and width stretching  from $\phi_{1} = -11^{\circ}$ to $\phi_{1} = -53^{\circ}$.

Here, we adopt the stream position and width from B18 for $\phi_{1}$ between $-10^{\circ}$ and $-54^{\circ}$.
Outside the region, we adopt the stream position of K10, assuming a width of $1^{\circ}$.
Finally, for the stream parts not covered by B18 and by K10, i.e. $-88^{\circ} \leq \phi_{1} < -62^{\circ}$, we adopt the stream position from W09, again assuming a width of $1^{\circ}$.
A summary of the stream positions adopted here is presented in Table\,1.

With the stream position and width adopt above, we then select stars observed by  either SDSS/SEGUE or LAMOST (with SNRs greater than 10) in the stream area.
The results are shown in Fig.\,1.
A total of 39,363 and 24,777 stars are selected in this way from the LAMOST and SDSS/SEGUE surveys, respectively.
The previous studies show that  the metallicity of GD-1 stream is at least poorer than $-1.4$.
We thus apply a metallicity cut of [Fe/H]\,$\leq -1.2$ to reduce the contamination of the thin and thick disk stars  by a significant amount.
The cut reduces the number of stars observed by the LAMOST and by the SDSS/SEGUE to 641 and 1411, respectively.
Combined together, this yields 1958 unique GD-1 member candidates.
For stars observed by both SDSS and LAMOST, we adopt the parameters yielded by the spectra of higher SNRs.

\begin{figure}
\begin{center}
\includegraphics[scale=0.55,angle=0]{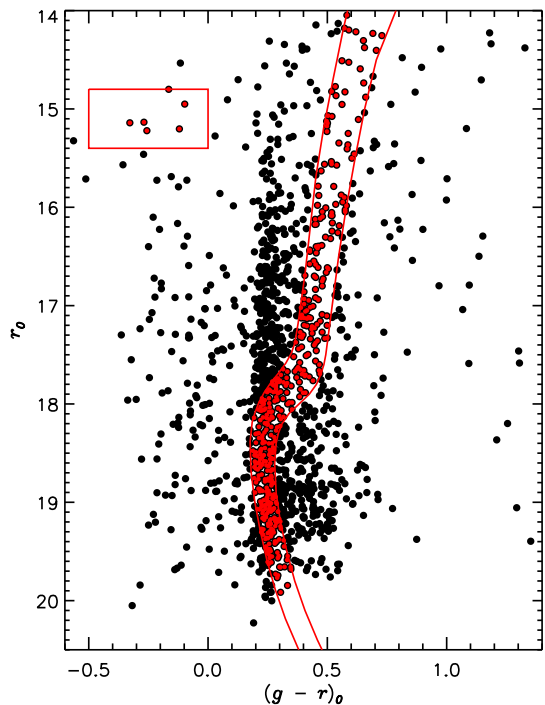}
\caption{The color-magnitude diagram (CMD) of the GD-1 stream of the 1958 stars, selected around the stream track (see Section\,3.1 for details) and corrected for the distance variations (see Section\,2 for details).
              The left and right lines represent the isochrones of M92 (with a shift of $-0.025$ in color) and M5 (with a shift of $+0.025$ in color), respectively.
              The red box is used to select BHBs of the GD-1 stream.
              The red dots mark the final selected stars of the GD-1 stream in the CMD.}
\end{center}
\end{figure}

\begin{table}
\centering
\caption{Position and width of the GD-1 stream}
\begin{threeparttable}
\begin{tabular}{ccc}
\hline
$\phi_{1}$& Width &Reference\\
(degree)&(degree)&\\
\hline
$[-10$, $+6]$&$\pm 1.0$&K10\\
$[-54$, $-10]$&$\pm 3\sigma$\tnote{\it a}&B18\\
$[-62$, $-54]$&$\pm 1.0$&K10\\
$[-88$, $-62]$&$\pm 1.0$&W09\\
\hline
\end{tabular}
\begin{tablenotes}
\item[$a$] Here $\sigma$ is the stream width (dispersion) listed in Table\,2 of B18.
\end{tablenotes}
\end{threeparttable}
\end{table}

\begin{table*}
\centering
\caption{Fitting results for Regions A to H}
\begin{threeparttable}
\begin{tabular}{cccccccccccc}
\hline
Region &$\phi_{1}$\,range& $\phi_{1}$\tnote{$a$} & $\phi_{2}$\tnote{$a$} & $f_{\rm st}$& $v_{\rm gsr}^{\rm st}$& $\sigma_{\rm gsr}^{\rm st}$ & $v_{\rm gsr}^{\rm MW}$& $\sigma_{\rm gsr}^{\rm MW}$ & $v_{\rm gsr}^{\rm BS}$\tnote{$b$}& $\sigma_{\rm gsr}^{\rm BS}$\tnote{$b$}&$N$\tnote{$c$} \\
&(degree) & (degree) & (degree) & & (km\,s$^{-1}$)&(km\,s$^{-1}$)&  (km\,s$^{-1}$)&(km\,s$^{-1}$)&  (km\,s$^{-1}$)&(km\,s$^{-1}$)&\\
\hline
  A &[$-85.0$, $-75.0$]&$ -81.67$&$  -1.64$&$0.29^{+0.19}_{-0.16}$&$  106.11^{+8.15}_{-12.14}$&$12.18^{+5.25}_{-6.73}$&$-40.45^{+55.91}_{-57.15}$&$134.20^{+34.35}_{-24.44}$&$-28.18$&$137.16$&$    3$\\
  B &[$-75.0$, $-62.5$]&$ -69.24$&$  -1.13$&$0.09^{+0.09}_{-0.06}$&$41.05^{+10.93}_{-8.31}$&$11.34^{+6.32}_{-8.64}$&$9.49^{+22.27}_{-22.65}$&$119.85^{+17.94}_{-12.34}$&$-2.10$&$107.46$&$   11$\\
  C &[$-62.5$, $-52.5$]&$ -55.71$&$  -0.57$&$0.15^{+0.11}_{-0.09}$&$0.33^{+14.45}_{-8.35}$&$ 13.57^{+4.69}_{-6.20}$&$19.45^{21.21}_{-20.56}$&$116.66^{+16.31}_{-10.68}$&$24.19$&$94.56$&$    8$\\
  D &[$-52.5$, $-38.0$]&$ -46.75$&$  +0.07$&$0.24^{+0.07}_{-0.07}$&$-3.68^{+2.44}_{-2.60}$&$ 7.22^{+4.32}_{-2.33}$&$-2.37^{+17.43}_{-17.06}$&$119.42^{+14.02}_{-12.13}$&$29.18$&$108.10$&$   25$\\
  E &[$-38.0$, $-25.5$]&$ -28.89$&$  -0.02$ &$0.33^{+0.09}_{-0.08}$&$-60.86^{+3.40}_{-3.56}$&$11.62^{+3.89}_{-3.14}$&$-18.38^{+18.10}_{17.96}$&$107.46^{+14.72}_{-11.78}$&$-23.27$&$108.04$&$   31$\\
  F &[$-25.5$, $-15.0$]&$ -23.29$&$  +0.10$&$0.41^{+0.10}_{-0.10}$&$-82.19^{+3.52}_{-3.32}$&$10.62^{+3.32}_{-2.79}$&$-1.47^{+28.15}_{-27.12}$&$115.12^{+22.50}_{-17.77}$&$-37.69$&$89.32$&$   18$\\
  G &[$-15.0$, $-5.0$]&$  -9.73$&$  -0.62$   &$0.17^{+0.10}_{-0.09}$&$-117.12^{+9.39}_{-11.65}$&$14.93^{+3.59}_{-5.35}$&$22.58^{+21.48}_{-22.14}$&$94.50^{+13.97}_{-9.75}$&$-20.42$&$118.57$&$    10$\\
  H &[$-5.0$, $5.0$]&$  -2.22$&$  -0.87$      &$0.33^{+0.11}_{-0.10}$&$ -130.37^{+5.64}_{-5.37}$&$12.91^{+4.18}_{-3.49}$&$29.56^{+31.29}_{-31.49}$&$108.58^{+24.19}_{-17.64}$&$1.28$&$109.61$&$    9$\\
\hline
\end{tabular}
\begin{tablenotes}
\item[$a$] These values are the median of the member candidates after applying the radial velocity cut.
\item[$b$] Means and dispersions of the radial velocity distributions (after the same cuts described in Sections\,3.1 and 3.2), obtained from the Besan{\c c}on model (Robin et al. 2003). 
\item[$c$] The number of member candidates in each region after applying the radial velocity cut.
\end{tablenotes}
\end{threeparttable}
\end{table*}

\begin{figure*}
\begin{center}
\includegraphics[scale=0.6,angle=0]{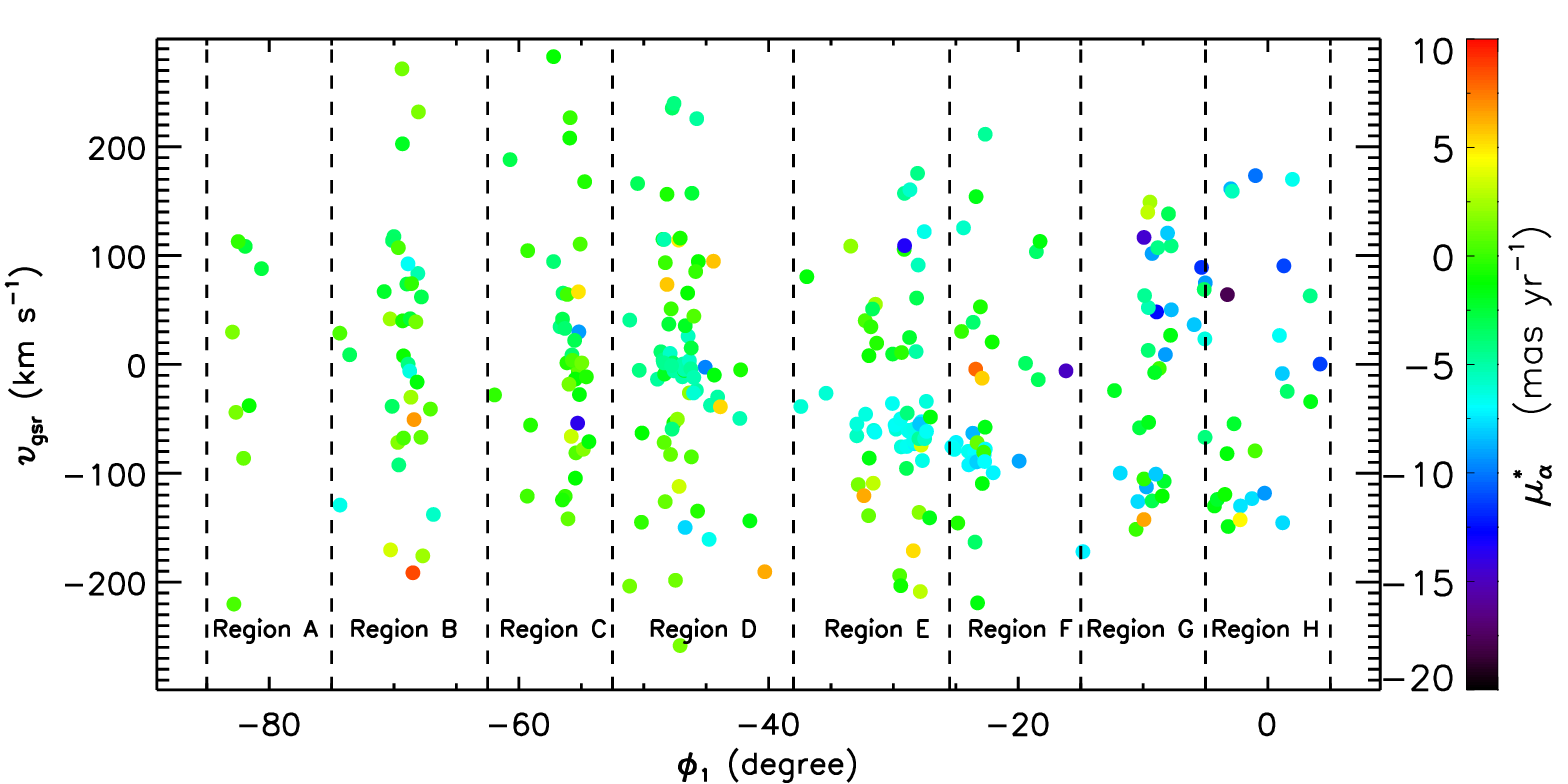}
\includegraphics[scale=0.6,angle=0]{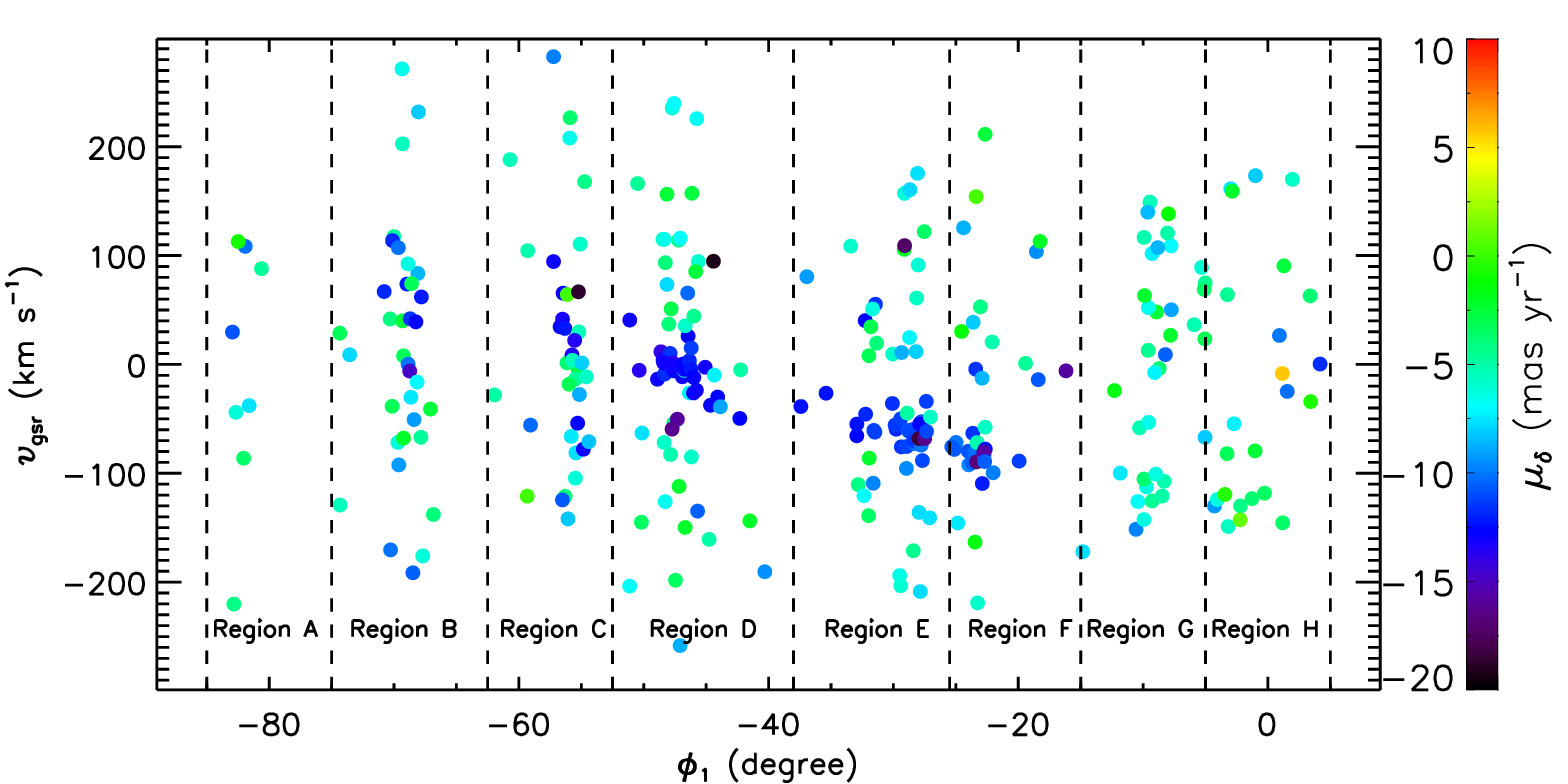}
\caption{Postion ($\phi_{1}$)--radial velocity ($v_{\rm gsr}$) diagram of the 307 GD-1 member candidates selected by applyimg the position, metallicity and CMD cuts.
              The data points are color-coded by proper motions in $\alpha$ (top panel) and $\delta$ (bottom panel).}
\end{center}
\end{figure*}

\begin{figure*}
\begin{center}
\includegraphics[scale=0.4,angle=0]{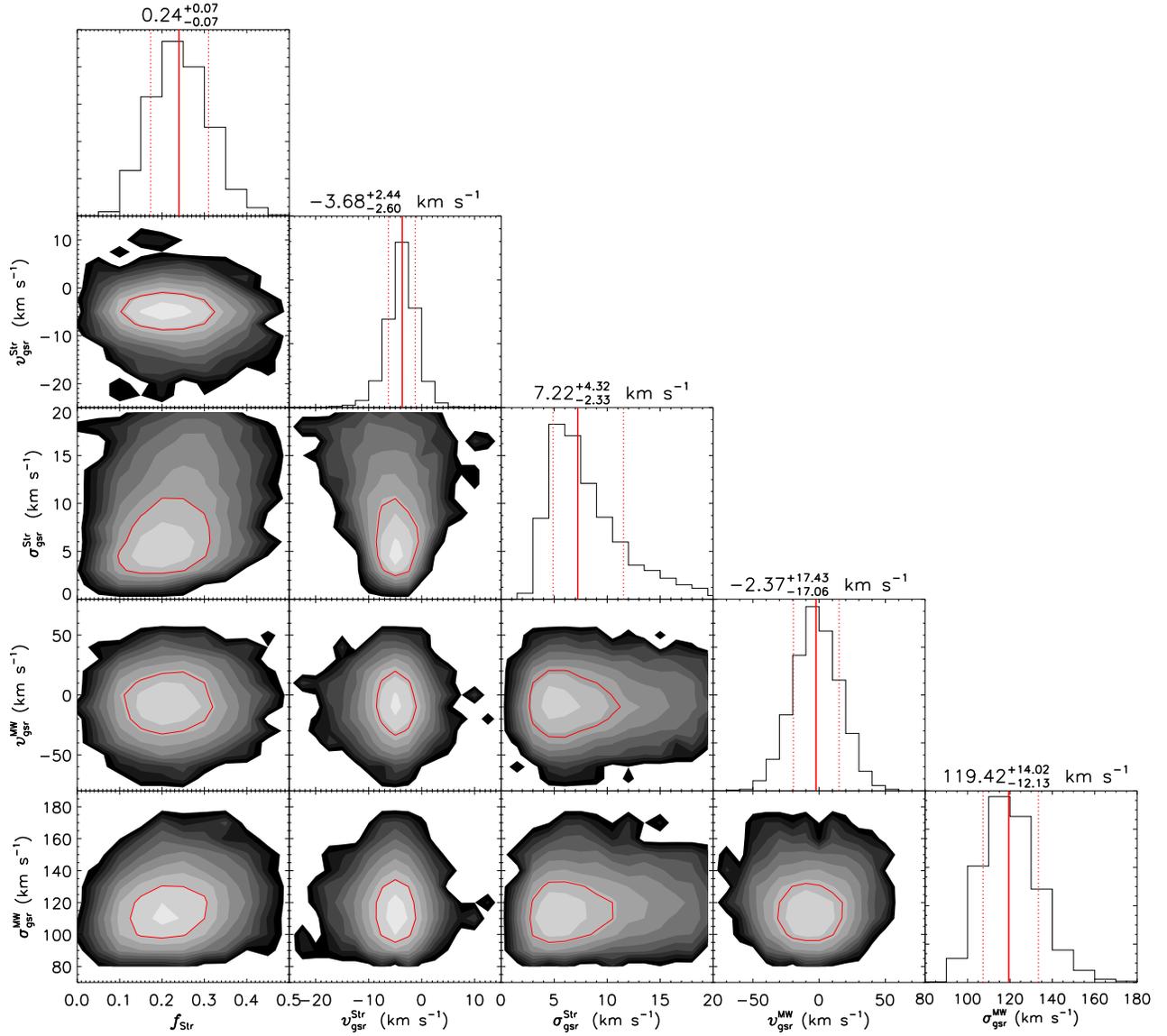}
\caption{Two-dimensional marginalized posterior probability distributions of the 5 assumed model parameters (see details in Section\,3.3) obtained with the MCMC technique. 
              The histograms overplotted on top of the individual columns present the one-dimensional marginalized posterior probability distributions of the parameters labelled at the bottom of the columns.
              The red contour in each panel indicates the $1\sigma$ confidence level.
              The red solid and dotted lines in each histogram represent the best-fit value and the 68 per cent probability intervals of the parameter concerned, respectively.
              The resulting values and uncertainties of the model parameters are also marked on the top of the individual columns.}
\end{center}
\end{figure*}

\begin{figure*}
\begin{center}
\includegraphics[scale=0.5,angle=0]{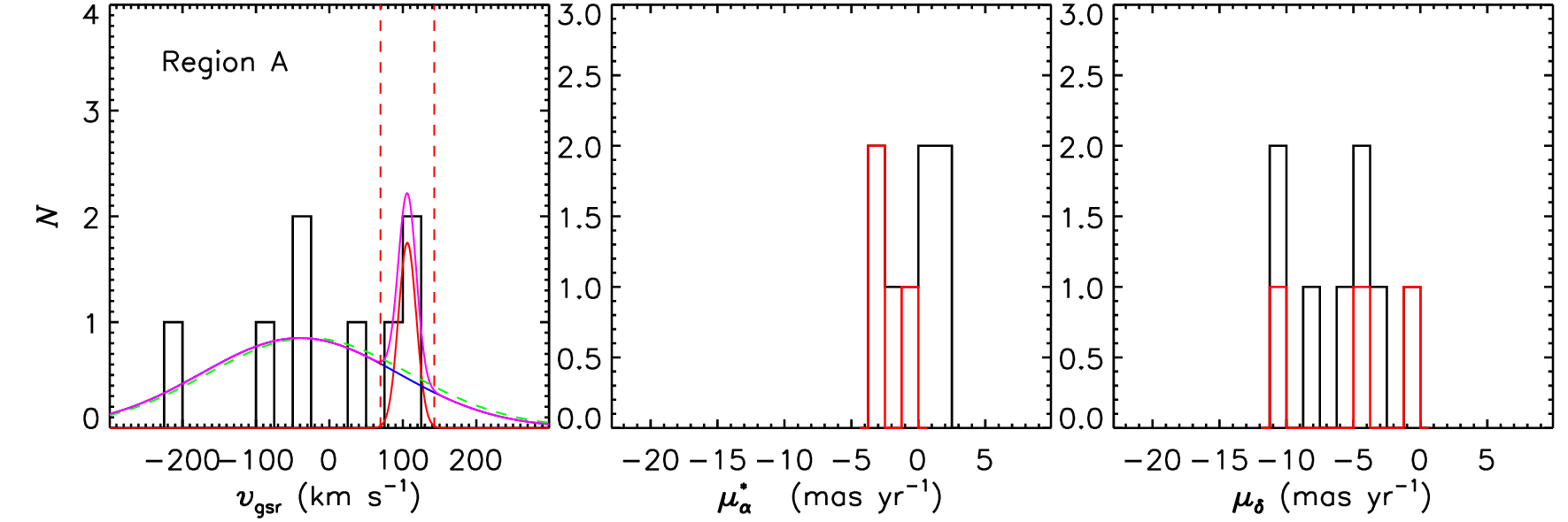}
\includegraphics[scale=0.5,angle=0]{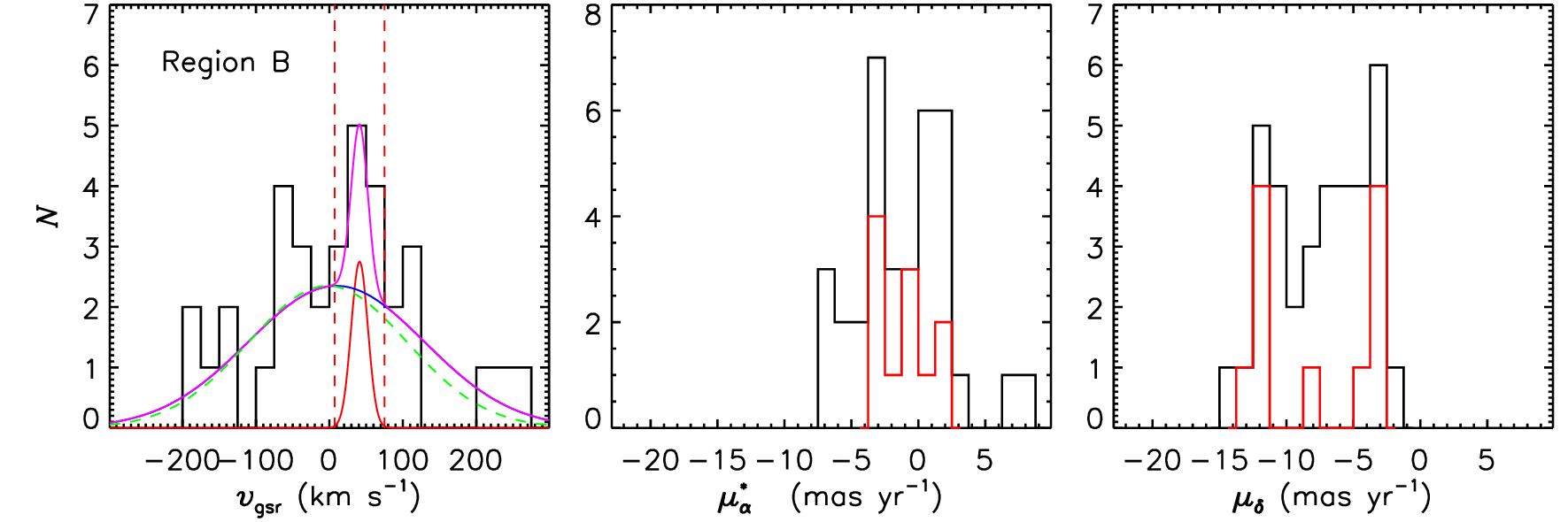}
\includegraphics[scale=0.5,angle=0]{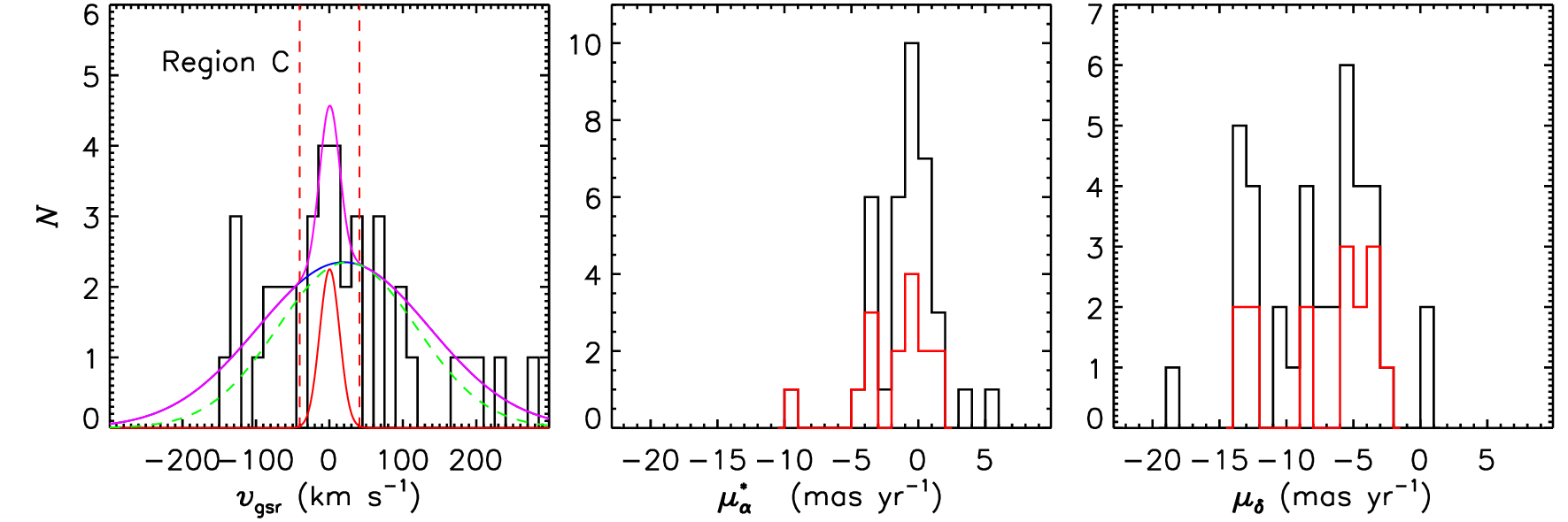}
\includegraphics[scale=0.5,angle=0]{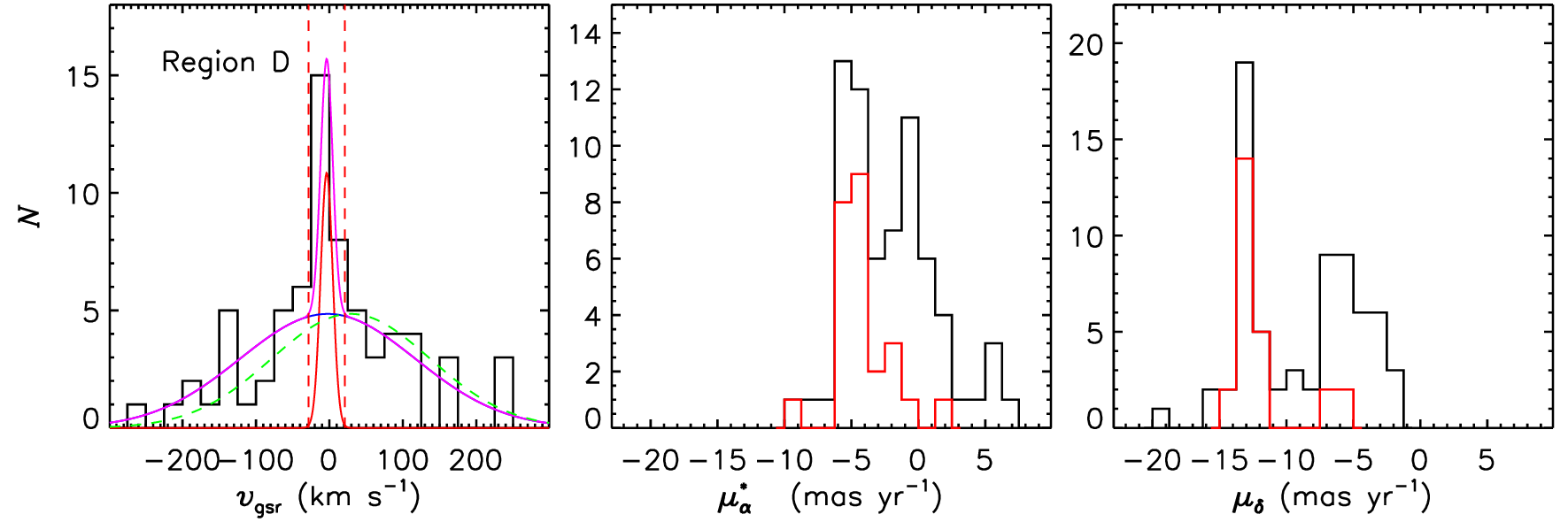}
\caption{Radial velocity and proper motion distributions for Regions A to D.
              The left panels show the radial velocity ($v_{\rm gsr}$) distributions of the different regions.
              The magenta lines in the individual panels represent the best-fit radial velocity distributions (arbitrary vertical scale; see Section\,3.3 for details).
              The red and blue lines mark the contributions of the stream and the background/foreground MW stars, respectively.
              The dashed green lines represent the radial velocity distributions (arbitrary vertical scale) from the simulated foreground/background MW stars, obtained from the Besan{\c c}on model (Robin et al. 2003). 
              The red dashed lines represent the radial velocity cut (i.e. $|v_{\rm gsr} - v_{\rm gsr}^{\rm st}| \leq 3\sigma_{\rm gsr}^{\rm st}$) used to select more probable candidates.
              The middle and right panels show the number distributions of proper motions in the $\alpha$ and $\delta$, respectively.
              The red histograms in both panels show the  distributions after applying the radial velocity cuts.}
\end{center}
\end{figure*}

\begin{figure*}
\begin{center}
\includegraphics[scale=0.5,angle=0]{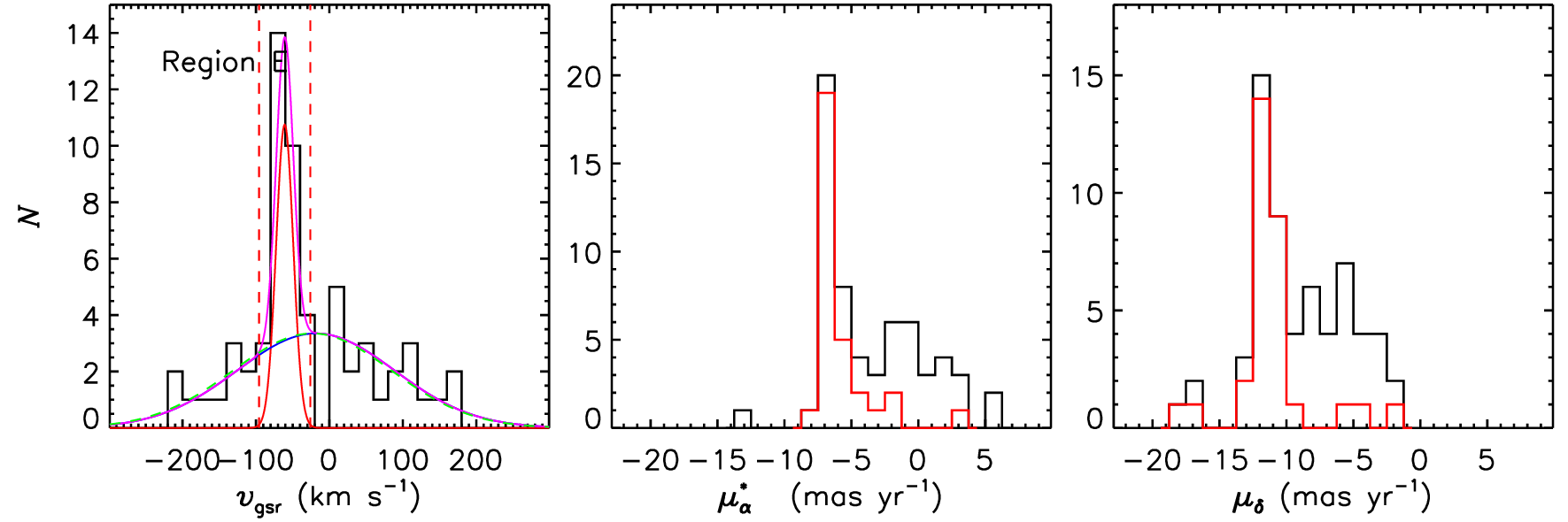}
\includegraphics[scale=0.5,angle=0]{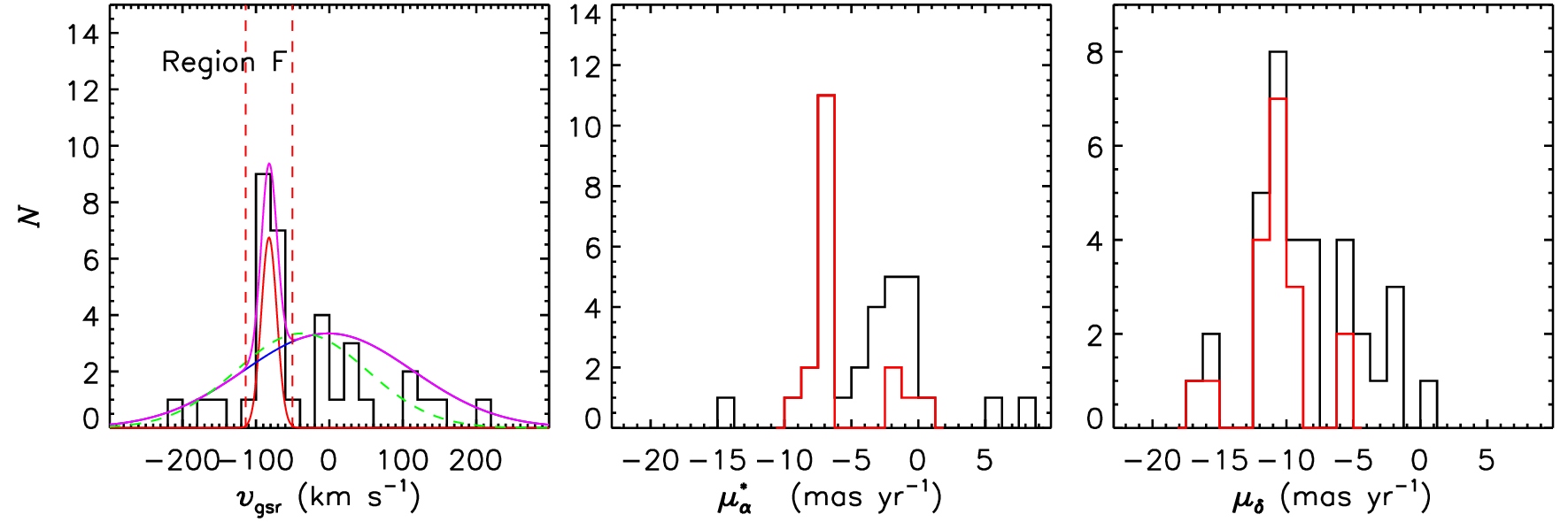}
\includegraphics[scale=0.5,angle=0]{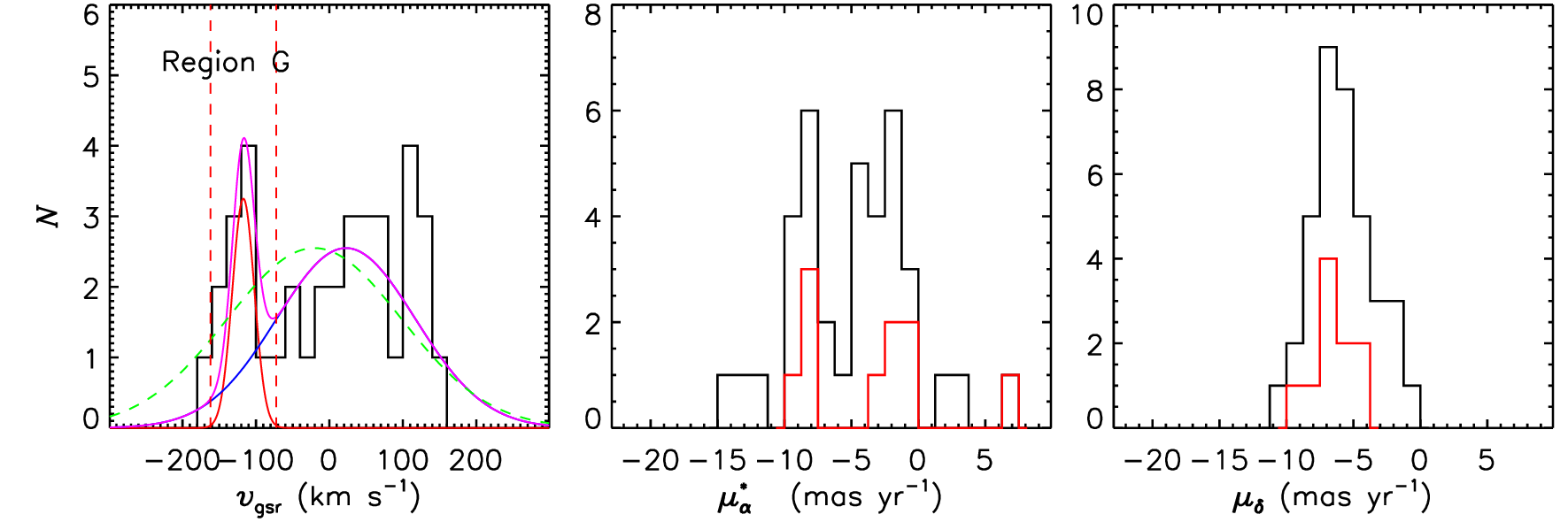}
\includegraphics[scale=0.5,angle=0]{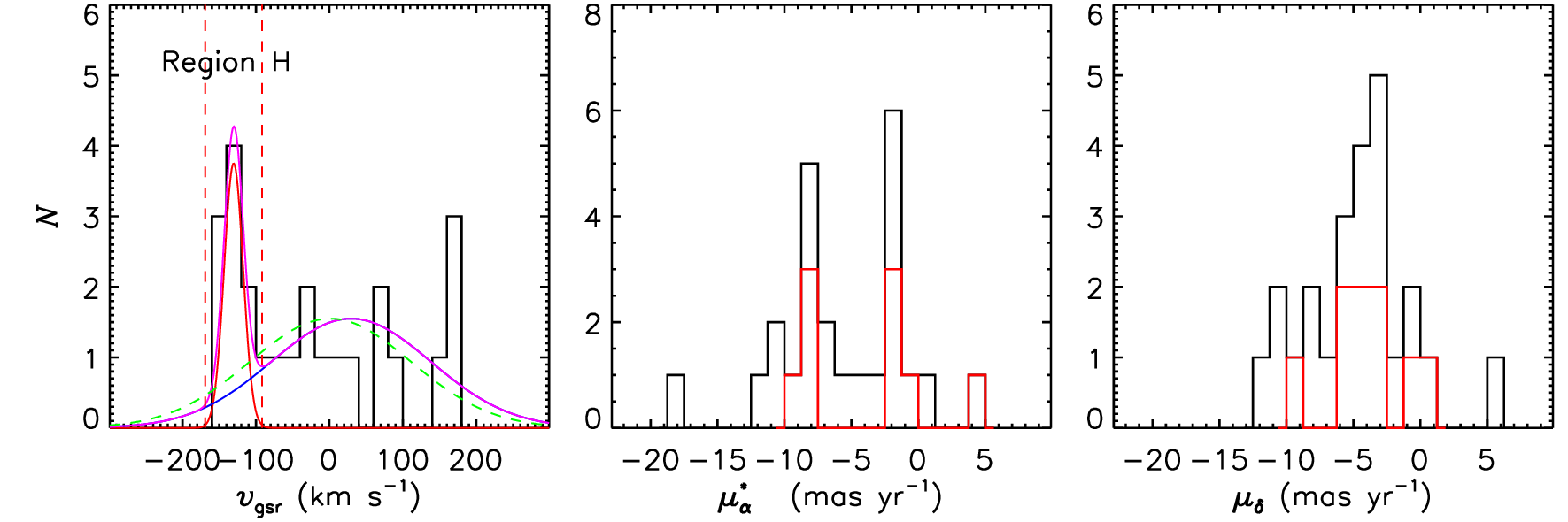}
\caption{Same as Fig.\,5 but for Regions E to H.}
\end{center}
\end{figure*}

\begin{figure*}
\begin{center}
\includegraphics[scale=0.6,angle=0]{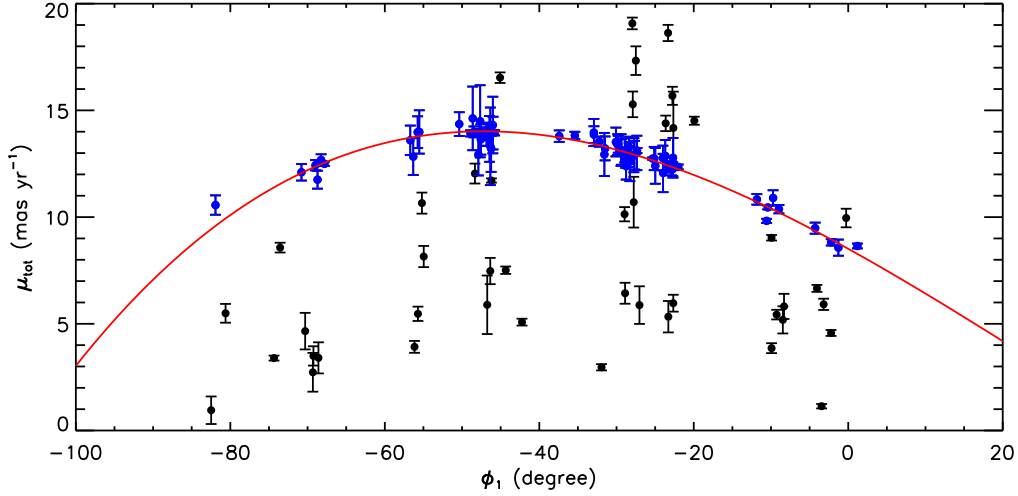}
\caption{Position ($\phi_{1}$)--proper motion ($\mu_{\rm tot}$) of the 115 member candidates of GD-1, selected from the position, metallicity, CMD and radial velocity cuts.
              The blue dots represent stars within 3$\sigma$ of the third-order polynomial fit (red line).}
\end{center}
\end{figure*}

\begin{figure*}
\begin{center}
\includegraphics[scale=0.6,angle=0]{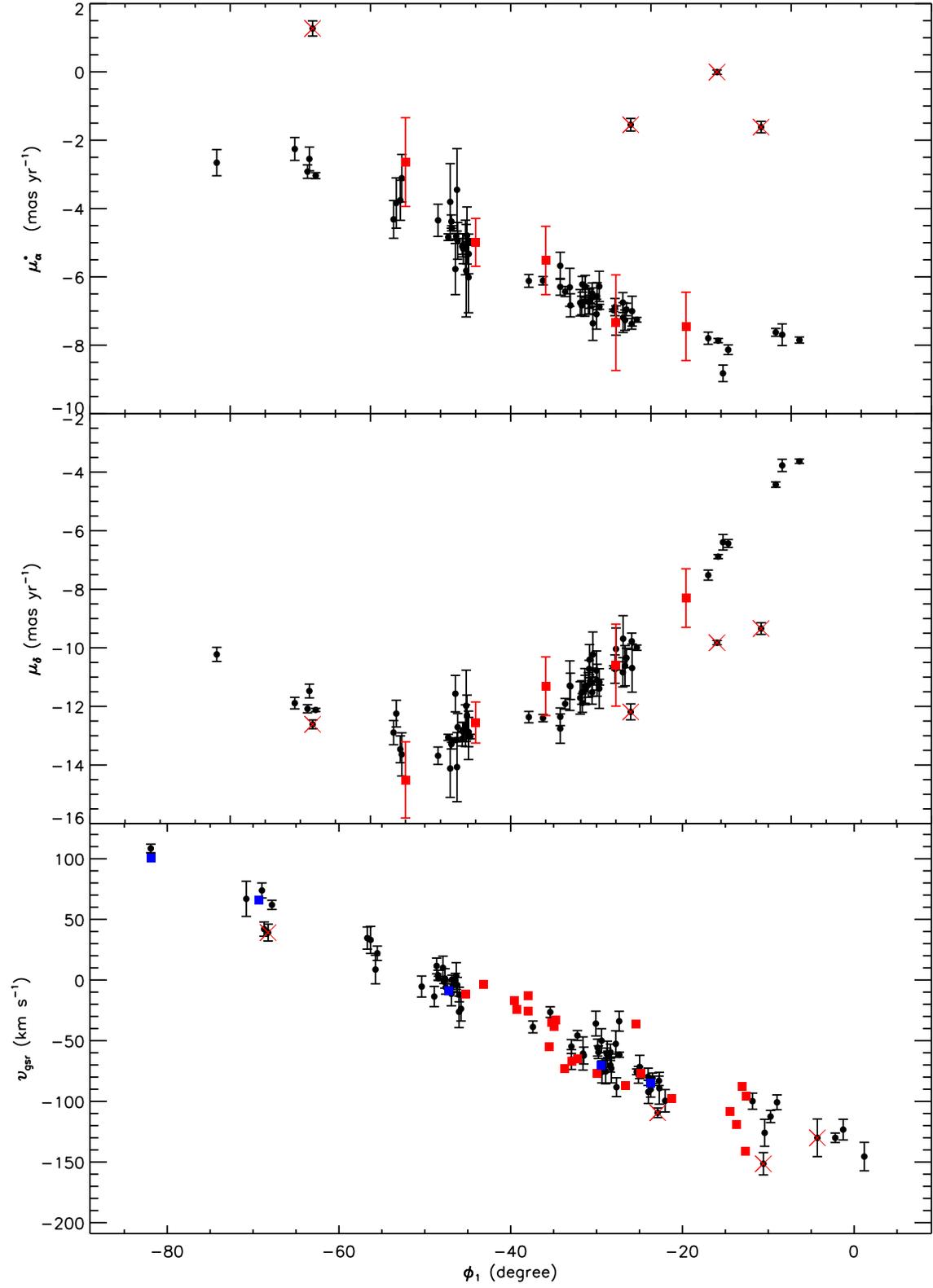}
\caption{Postion ($\phi_{1}$)--$\mu_{\alpha}^{*}$ (top panel), position ($\phi_{1}$)--$\mu_{\delta}$ (middle panel) and position ($\phi_{1}$)--$v_{\rm gsr}$ diagrams of the 71 member candidates of GD-1.
              The red crosses mark the three contaminators, identified in the $\phi_{1}$--$\mu_{\alpha}^{*}$ diagram.
              The blue and red squares are the results from W09 and K10, respectively.}
\end{center}
\end{figure*}

\begin{figure}
\begin{center}
\includegraphics[scale=0.55,angle=0]{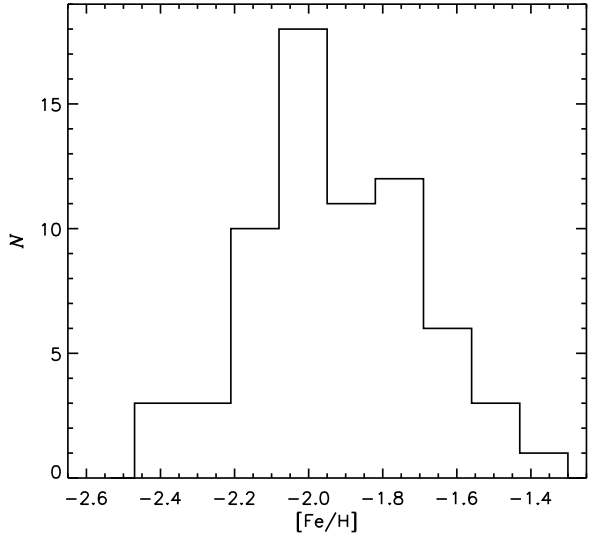}
\caption{Metallicity distribution of the 67 final highly probable member stars of the GD-1 stream.}
\end{center}
\end{figure}

\begin{figure*}
\begin{center}
\includegraphics[scale=0.55,angle=0]{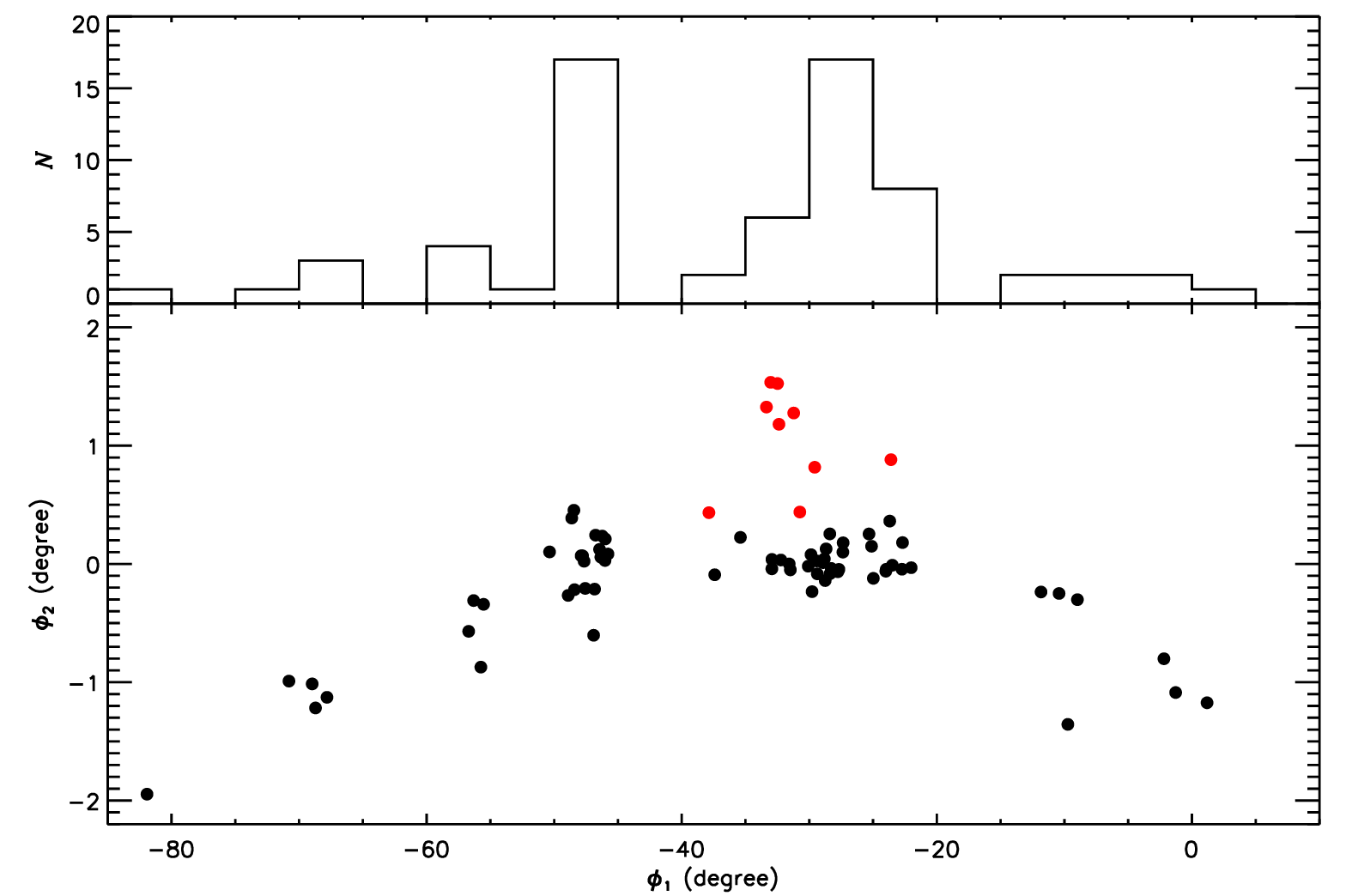}
\caption{Spatial distribution of the final 67 stream candidates and 9 member candidates of a newly detected fanning structure (see Section\,4 for details).
               The histogram overplotted on top shows the number distribution of those stream candidates along $\phi_1$.}
\end{center}
\end{figure*}

\subsection{Color-magnitude diagram}
From the 1958 candidate stars selected by position and metallicity cuts, we further identify those more probable ones in the color-magnitude diagram (CMD).
Fig.\,2 shows $(g-r)_{0}$ versus $r_{0}$ diagram of the 1958 stars.
here we have corrected for variations in the distance of the stream, by shifting the magnitudes of stars at different positions of the stream to a common distance modulus of $(m-M)_{0} = 14.58$.
The distance modulus as a function of position is described by a second-order polynomial,  $f(\phi_{1}) = 14.58 + 2.923\times10^{-4}(\phi_{1}+44.66)^{2}$, derived by B18.
Values of the interstellar reddening of the individual stars are adopted from Schlegel, Finkbeiner \& Davis (1998).
Then we use two empirical isochrones of globular clusters to single out more probable member candidates of the GD-1 stream.
They are M5 of a distance modulus of $14.26$ (An et al. 2009) and a metallicity of [Fe/H]\,=\,$-1.26$ (Kraft \& Ivans 2004), and M92 of a distance modulus of $14.60$ (Harris et al. 2010) and a metallicity of [Fe/H]\,=\,$-2.42$ (Kraft \& Ivans 2004).
The metallicities of these two clusters happen to bracket the most probable metallicity of the stream.
To allow for the potential uncertainties in photometry, we shift the isochrones of M92 and M5 by $-0.025$ and $0.025$\,mag in $(g-r)_{0}$, respectively.
In addition to the main isochrones, we also add a box of $-0.5 \leq (g-r)_{0} \leq 0.0$ and $14.80 \leq r_{0} \leq 15.40$ to select probable blue horizontal branch stars (BHBs) of the GD-1 stream.
With the above isochrones and box, 307 stars are selected.

\subsection{Radial velocities and proper motions}
In Subsections 3.1\,\&\,3.2, we have obtained 307 member candidates of the GD-1 stream, selected based on the stream position (Fig.\,1), metallicity and CMD  (Fig.\,2).
In this Subsection, we further reject potential contamination from the halo populations using the kinematic information (i.e. radial velocities and proper motions) to single out the most probable members of the GD-1 stream.
Doing so, we first show that the position($\phi_{1}$)-radial velocity ($v_{\rm gsr}$) diagram color coded by proper motions in Fig.\,3 and a clean stripe is clearly seen in this diagram (at least for $\phi_{1} \geq -55^{\circ}$, i.e. Regions D to H).
Here $v_{\rm gsr}$ is radial velocity in the Galactic standard of rest (GSR) frame given by:
\begin{equation}
v_{\rm gsr} = v_{\rm los} + U_{\odot}\cos(b)\cos(l) + v_{\phi,\odot}\cos(b)\sin(l) + W_{\odot}\sin{b}\text{,}
\end{equation}
where the values of the solar motion in the radial and vertical directions are adopted from Huang et al. (2015), i.e., ($U_{\odot}$, $W_{\odot}$) = (7.01, 4.95) km\,s$^{-1}$.
The value of $v_{\phi,\odot}$, i.e. $V_{\rm c} (R_{0})$ + $V_{\odot}$, is set to the value yielded by the proper motions of Sgr\,A$^{*}$ (Reid \& Brunthaler 2004) and $R_{0} = 8.34$\,kpc (Reid et al. 2014).
The proper motions (especially in $\delta$ direction) of this stripe are also significantly different from the field stars outside.
Therefore, stars on this stripe are the most probable members of the GD-1 stream. 

To select stars on the stripe, we divide them into eight regions in $\phi_{1}$ (i.e. Regions\,A to H).
For each region, we assume that the radial velocity distribution is a combination of true stream stars and foreground/background MW stars.
For a given region, the radial velocities of stream stars are assumed to follow a Gaussian distribution with a mean $v_{\rm gsr}^{\rm st}$ and a dispersion $\sigma_{\rm gsr}^{\rm st}$.
The number fraction of stream stars is set to $f_{\rm st}$.
As a result of the metallicity and CMD cuts, the foreground/background contaminators are largely from the halo population and thus their radial velocity distribution also can be represented by a Gaussian distribution characterized by a mean $v_{\rm gsr}^{\rm MW}$ and a dispersion $\sigma_{\rm gsr}^{\rm MW}$.
In this way, for the $i$th sample star of model parameters $\Theta =$\,\{$v_{\rm gsr}^{\rm st}$,\,$\sigma_{\rm gsr}^{\rm st}$,\,$f_{\rm st}$,\,$v_{\rm gsr}^{\rm MW}$,\,$\sigma_{\rm gsr}^{\rm MW}$\}, the likelihood of observing the star to have a radial velocity $v_{\rm gsr}$ is given by,
\begin{equation}
L_{i} (v_{\rm gsr}^{i}|\Theta) = f_{\rm st}P_{\rm st} (v_{\rm gsr}^{i}) + (1 - f_{\rm st})P_{\rm MW}(v_{\rm gsr}^{i})\text{.} 
\end{equation}

$P_{\rm st} (v_{\rm gsr}^{i})$ is the probability of a stream star having a measured radial velocity $v_{\rm gsr}^{i}$ and can be calculated from,
\begin{equation}
P_{\rm st} (v_{\rm gsr}^{i}) = \frac{1}{\sqrt{2\pi}\sigma_{\rm gsr}^{\rm st}}\exp{[-\frac{1}{2}(\frac{v_{\rm gsr}^i-v_{\rm gsr}^{\rm st}}{\sigma_{\rm gsr}^{\rm st}})^2]}\text{.} 
\end{equation}

$P_{\rm MW} (v_{\rm gsr}^{i})$ is the probability of a MW foreground/background star having a measured radial velocity $v_{\rm gsr}^{i}$ and can be obtained from,
\begin{equation}
P_{\rm MW} (v_{\rm gsr}^{i}) = \frac{1}{\sqrt{2\pi}\sigma_{\rm gsr}^{\rm MW}}\exp{[-\frac{1}{2}(\frac{v_{\rm gsr}^i-v_{\rm gsr}^{\rm MW}}{\sigma_{\rm gsr}^{\rm MW}})^2]}\text{.} 
\end{equation}

Multiplying the expression of Eq.\,(2) for a total of $N_j$ stars for a specific region $j$, the likelihood of this region is then given by,
\begin{equation}
L = \prod_{i=1}^{N_j} L_i\text{.} 
\end{equation}

The posterior distributions of the assumed parameters of our model can be derived from,
\begin{equation}
p (\Theta|{\bold O}) \propto L({\bold O}|\Theta)I(\Theta)\text{,}
\end{equation}
where \textbf{O} represents the observables, i.e. \{$v_{\rm gsr}^{i}$\}$_{i=1}^{N_j}$.
$I (\Theta)$ represents the priors of the assumed parameters.
Details of those priors for the individual regions are listed in Appendix\,A.
To obtain the posterior probability distributions of those assumed parameters, the Markov Chain Monte Carlo (MCMC) technique is applied for each region (i.e. Regions A to H). 
As an example, the resulted posterior probability distributions of the five free model parameters for Region\,D are shown in Fig.\,4.
Generally, those joint distributions indicate that those free parameters are largely independent to each other.
The best-fit values and uncertainties of those model parameters are estimated by the median values and the 68 per cent probability intervals  of their marginalized posterior probability distributions, respectively.
All the best-fit results are shown in Figs.\,5 and 6, and Table\,2.
As shown in Figs.\,5 and 6, the fitted foreground/background MW radial velocity distributions are generally consistent with the simulated ones as given by the Besan{\c c}on model (Robin et al. 2003).
Finally, we note the following points:
1) The estimated $\sigma_{\rm gsr}^{\rm st}$ does not represent the intrinsic radial velocity dispersion of the stream stars, which is actually the quadratic sum of the typical radial velocity uncertainty (about 5-10 km\,s$^{-1}$) and the intrinsic radial velocity dispersion; 
2) The method of fitting the radial velocity distribution in order to further single out the stream stars is a traditional technique used in many previous studies (e.g. Kopsov, Rix \& Hogg 2010; Ishigaki et al. 2016);
3) For Regions A to C, the prior assumption on $v_{\rm gsr}^{\rm st}$ (see Appendix\,A) is crucial for estimating the parameters described above, due to the high fractions of the foreground/background MW stars and the heavily overlapping between the radial velocity distributions of the stream and of the field stars in those three regions.

To illustrate the validity of our model selection in fitting the observed radial velocity distributions of different regions, we repeat all the above analysis by assuming a purely halo population, i.e. just one Gaussian distribution without stream component.
The values of Bayesian Information Criterion (BIC; Schwarz 1978) of the two models at different regions are calculated and presented in Table\,3.
As expected, in Regions A to C (and G), the values of BIC of the two models are nearly the same, indicating no differences between the two models. 
In other regions, the values of BIC of the model with stream component is significantly smaller than those of the model without stream component, and thus the model with stream component is preferred in those regions.

 \begin{table*}
\centering
\caption{Values of  Bayesian Information Criterion of different models}
\begin{tabular}{cccccccccc}
\hline
Model&A&B&C&D&E&F&G&H\\
\hline
Stream + Halo populations&100.09&435.30&471.24& 829.33& 714.29& 407.98&  450.90&   292.42\\
Halo population&101.67& 434.01& 465.54& 841.61& 726.53& 422.60& 450.02&   299.94\\
\hline
\end{tabular}
\end{table*}

With the model parameters estimated above (see Table\,2), we then select the member stars of the GD-1 stream in a specific region as those of radial velocities that satisfy $|v_{\rm gsr} - v_{\rm gsr}^{\rm Str}| \leq 3\sigma_{\rm gsr}^{\rm Str}$ in that region. 
With this cut, a total of 115 stars are selected in Regions A to H.
The potential contaminators are further rejected in the position ($\phi_{1}$)--proper motion ($\mu_{\rm tot}$)  diagram.
Doing this, we show the $\phi_{1}$--$\mu_{\rm tot}$ plot in Fig.\,7 and a clear stripe is seen with some significant outliers (i.e. contaminators).
We fit this stripe with a third-order polynomial.
We iterate the fitting, discarding data points that deviate more than $3\sigma$ from the fit.
Finally, we obtain 71 highly probable member stars of the GD-1 stream.

 \begin{table*}
\centering
\caption{Properties of the final 67 highly probable member stars of GD-1 stream}
\begin{tabular}{cccccccc}
\hline
Name& $G$&$g$ & $r$& $v_{\rm los}$ & [Fe/H] & $\mu_{\alpha}$&$\mu_{\delta}$ \\
& (mag) & (mag)& (mag) & (km\,s$^{-1}$)&&(mas\,yr$^{-1}$)&(mas\,yr$^{-1}$)\\
\hline
J08:26:34.90-00:20:25.8&$17.983 \pm 0.002$&$18.296 \pm 0.022$ & $17.967 \pm 0.006$&$267.32 \pm 3.56$&$-1.62$&$-2.658 \pm 0.384$& $ -10.224  \pm 0.241$\\
...&...&...&...&...&...&...&...\\
\hline
\end{tabular}
{Note: This table is available online in its entirety in machine-readable form.}
\end{table*}

\subsection{The final stellar stream member stars}
In Fig.\,8, we show $\phi_{1}$--$\mu_{\alpha}^{*}$, $\phi_{1}$--$\mu_{\delta}$ and $\phi_{1}$--$v_{\rm gsr}$ diagrams of the 71 highly probable member stars as selected above. 
In the plot of $\phi_{1}$--$\mu_{\alpha}^{*}$, there are four obvious outliers that escape all the cuts applied and sneak into the sample.
As the plots show, the proper motions in the $\alpha$ direction are about $-3$\,mas\,yr$^{-1}$ at $\phi_{1} = -80^{\circ}$ and then almost linearly decrease to $-8$\,mas\,yr$^{-1}$ at $\phi_{1}$ around $-5^{\circ}$.
The proper motions in the $\delta$ direction show $-10$\,mas\,yr$^{-1}$ at $\phi_{1} = -80^{\circ}$ and decrease to a minimum value of about $-13$\,mas\,yr$^{-1}$ at $\phi_{1}$ around $-50^{\circ}$, and then go up to $-4$ \,mas\,yr$^{-1}$ at $\phi_{1} = 0^{\circ}$. 
As a comparison, the proper motions of GD-1 found by K10 between $-55^{\circ}$ and $-15^{\circ}$ in $\phi_1$ are also shown in Fig.\,8.
Except some small systematic offsets in $\phi_1 = -35^{\circ}$ and $-55^{\circ}$ (still within the typical uncertainties), their results are generally consistent with ours.
The radial velocities show almost a monotone decrease, from $100$\,km\,s$^{-1}$ at $\phi_{1} = -80^{\circ}$ to $-150$\,km\,s$^{-1}$ at $\phi_{1} = 0^{\circ}$.
The metallicity distribution of the final 67 highly-probable GD-1 member stars is shown in Fig.\,9, with a median value of $-1.96$ and a dispersion of 0.23\,dex.
The median value is consistent with the previous studies (e.g. W09).

We note that in the position ($\phi_{1}$)--radial velocity ($v_{\rm gsr}$) diagram, the new data points now encompass almost the entire length of the stream.
In W09, the mean radial velocities are available only at five positions (see the bottom panel of Fig.\,8).
The trend of the stream in the position--radial velocity diagram as derived by W09 is in excellent agreement with what find here.
K10 obtain radial velocities of 21 highly probable member candidates, covering $\phi_{1}$ from $-45^{\circ}$ to $-12^{\circ}$.
Again, the trend of the stream in the position--radial velocity found by K10 is consistent with what we find here.

Finally, properties (name/coordinates, Gaia $G$ and SDSS $gr$ magnitudes, line-of-sight velocity, metallicity [Fe/H] and proper motions from Gaia DR2) of the 67 stars are presented in Table\,4.

\section{Discussions}
With the accurate spectroscopic (i.e. [Fe/H] and line-of-sight velocity) and astrometric (i.e. proper motions) properties of the selected  67 highly probable member stars of the GD-1, one can now explore the possible origin of this stream. 
Although there is no apparent progenitor around the stream, the low metallicity measured from the spectra and its thin, cold nature indicate the stream is likely originated from a disrupted globular cluster rather than a dwarf galaxy.
In principle, the dispersion of the metallicity distribution (see Fig.\,9) of our final stream candidates can provide vital constraints on the nature of the stream progenitor.
Unfortunately, the typical metallicity uncertainties  of those metal-poor stars derived from our low resolution spectra are quite large, about 0.2-0.3\,dex (see Section\,2), comparable to the dispersion (0.23\,dex) of the metallicity distribution as mentioned in Section\,3.4.

More recently, by studying the morphology and density distribution of GD-1 using the deep photometric data from CFHT/Megacam, B18 detect a significant under-density around $\phi_1 = - 45^{\circ}$, and suggest that the missing progenitor of this stream is probably at this location.
To confirm their results, we show the spatial distribution of our final stream candidates in Fig.\,10. 
The number distribution along $\phi_1$ clearly show an under-density around $\phi_1 = - 45^{\circ}$, with two peaks on either sides.
Our new result confirms B18's findings and supports interpretation, although our stream member stars have not selected from a completed sample and thus may suffer from potential selection effects.

Using the proper motions measured from the Pan-STARRS1 survey and a new stream detection algorithm -- {\it STREAMFINDER}, Malhan et al. (2018) have detected a new $\sim 40^{\circ}$ long structure approximately parallel to GD-1.
They propose that this new structure is possibly a sign of stream-fanning, caused by the triaxiality of the Galactic potential.
As expected, we do not see such a long structure in our final member candidates in Fig.\,10.
This is simply because that our member candidates are selected along the GD-1 track in the sky as provided by the previous studies (e.g. W09, K10 and B18; see Section\,3.1).
To confirm the existence of the fanning structure, we apply the metallicity, CMD, kinematic (radial velocity and proper motion) cuts described above to the off-stream stars (but with $\phi_2$ no smaller than $-2^{\circ}$ and no larger than $2.5^{\circ}$).
In total, 9 stars are left and largely distributed between $-40$ and $-20^{\circ}$ in $\phi_1$ with $\phi_2$ above the stream stars.
The positions of those 9 stars are consistent with the fanning structure found by Malhan et al. (2018), strongly supporting their new finding.
Given the limited number of the currently identified probable member stars of the newly detected, long fanning structure, future spectroscopic observations of more stars in those off-stream regions  are desirable.

Finally,  the rich information of the stream member stars can also be used to constrain the mass distribution and the gravitational potential of the Milky Way.
For this, we leave it to a future paper.

\section{Summary}
With photometric data from the SDSS survey, spectroscopic data from the SDSS/SEGUE and LAMOST surveys, and astrometric data from the Gaia DR2, a total of 67 highly-probable member stars of the GD-1 have been selected by applying a variety of cuts, in sky position, metallicity, CMD and in proper motions.
The newly identified member stars cover almost the entire length of the stream.
With the available spectroscopic and astrometric information of those member stars, we have obtained position-velocity diagrams, $\phi_{1}$--$\mu_{\alpha}$, $\phi_{1}$--$\mu_{\delta}$ and $\phi_{1}$--$v_{\rm gsr}$  planes of the stream.
In the $\phi_{1}$--$v_{\rm gsr}$ plane, our results are consistent with previous work, but have greatly extended the spatial extent, covering almost the entire length of the stream.
In addition, we obtain a mean metallicity of this stream of $-1.96$.

From the spatial distribution of the stream member stars, we confirm a under-density around $\phi_1 = - 45^{\circ}$ found by B18, which is a promising location of the elusive missing progenitor of the stream.
We also identify few member candidates of a long fanning structure (approximately parallel to GD-1) newly detected by Malhan et al. (2018).
Besides, the abundant information of the stream member stars presented in the current work can also place strong constraints on the mass distribution and the gravitational potential of the Milky Way.

 \section*{Acknowledgements} 
The Guoshoujing Telescope (the Large Sky Area Multi-Object Fiber Spectroscopic Telescope, LAMOST) is a National Major Scientific Project built by the Chinese Academy of Sciences. Funding for the project has been provided by the National Development and Reform Commission. LAMOST is operated and managed by the National Astronomical Observatories, Chinese Academy of Sciences.
The LAMOST FELLOWSHIP is supported by Special fund for Advanced Users, budgeted and administrated by Center for Astronomical Mega-Science, Chinese Academy of Sciences (CAMS).

This work has made use of data from the European Space Agency (ESA) mission Gaia (https://www.cosmos.esa.int/gaia), processed by the Gaia Data Processing and Analysis Consortium (DPAC, https://www.cosmos.esa.int/web/gaia/dpac/consortium).

This work is supported by the National Natural Science Foundation of China U1531244, 11833006, 11811530289, U1731108, 11473001 and 11803029.

\appendix
\section{Priors of the parameters of the radial velocity distribution model}

The priors of the assumed parameters of the radial velocity distribution model of each region are all uniform within the ranges listed in Table\,A1.

\begin{table}
\centering
\caption{Priors of the parameters of the radial velocity distribution model}
\begin{tabular}{cccccc}
\hline
Region & $f_{\rm st}$ & $v_{\rm gsr}^{\rm st}$ & $\sigma_{\rm gsr}^{\rm st}$ & $v_{\rm gsr}^{\rm MW}$ &$\sigma_{\rm gsr}^{\rm MW}$ \\
&&(km\,s$^{-1}$)&(km\,s$^{-1}$)&(km\,s$^{-1}$)&(km\,s$^{-1}$)\\
\hline
A&$[0, 1]$&$[50, 150]$&$[0, 20]$&$[-150, 150]$&$[80, 200]$\\
B&$[0, 1]$&$[20, 90]$&$[0, 20]$&$[-150, 150]$&$[80, 200]$\\
C&$[0, 1]$&$[-20, 70]$&$[0, 20]$&$[-150, 150]$&$[80, 200]$\\
D&$[0, 1]$&$[-40, 40]$&$[0, 20]$&$[-150, 150]$&$[80, 200]$\\
E&$[0, 1]$&$[-100, 50]$&$[0, 20]$&$[-150, 150]$&$[80, 200]$\\
F&$[0, 1]$&$[-100, 50]$&$[0, 20]$&$[-150, 150]$&$[80, 200]$\\
G&$[0, 1]$&$[-200, 50]$&$[0, 20]$&$[-150, 150]$&$[80, 200]$\\
F&$[0, 1]$&$[-200, 50]$&$[0, 20]$&$[-150, 150]$&$[80, 200]$\\
\hline
\end{tabular}
\end{table}

\end{document}